# Extraction of $n = 0$ pick-up by locked mode detectors based on neural networks in J-TEXT

Chengshuo Shen[1], Jianchao Li[2*], Yonghua Ding[1], Jiaolong Dong[1], Nengchao Wang[1], Dongliang Han[1], Feiyue Mao[1], Da Li[1], Zhipeng Chen[1], Zhoujun Yang[1], Zhongyong Chen[1], Yuan Pan[1] and J-TEXT team[†]

[1]International Joint Research Laboratory of Magnetic Confinement Fusion and Plasma Physics, State Key Laboratory of Advanced Electromagnetic Engineering and Technology, School of Electrical and Electronic Engineering, Huazhong University of Science and Technology, Wuhan, 430074, China

[2]Hubei Key Laboratory of Optical Information and Pattern Recognition, Wuhan Institute of Technology, Wuhan 430205, People's Republic of China

*Abstract* -- Measurement of locked mode (LM) is important for the physical research of Magnetohydrodynamic (MHD) instabilities and plasma disruption. The $n = 0$ pick-up need to be extracted and subtracted to calculate the amplitude and phase of the LM. A new method to extract this pick-up has been developed by predicting the $n = 0$ pick-up $b_r^{n=0}$ by the LM detectors based on Neural Networks (NNs) in J-TEXT. An approach called Power Multiple Time Scale (PMTS) has been developed with outstanding regressing effect in multiple frequency ranges. Three models have been progressed based on PMTS NNs. PMTS could fit the $b_r^{n=0}$ on the LM detectors with little errors both in time domain and frequency domain. The $n>0$ pick-up $b_r^{n>0}$ generated by resonant magnetic perturbations (RMPs) can be obtained after subtracting the extracted $b_r^{n=0}$. This new method uses only one LM instead of 4 LM detectors to extract $b_r^{n=0}$. Therefore, the distribution of the LM detectors can also be optimized based on this new method.

*Index Terms*-- J-TEXT Tokamak; Locked mode; Extraction of $b_r^{n=0}$; Neural Networks.

## I. Introduction

Tearing mode (TM) [1]-[3] is one of the most harmful Magnetohydrodynamics (MHD) instabilities even leading to disruption. The magnetic field generally has an island helical structure called magnetic island. The low-m tearing modes with $n=1$ or 2 can lead to a degradation of plasma confinement, such as the $m/n = 3/2$ and $2/1$ modes. If these kinds of modes stop rotation will become locked mode (LM), which can even cause major disruptions[4]-[5]. Therefore, measuring the $n=1$ or 2 LM is one necessary topic in MHD and disruption researches. Mirnov probes and saddle loops for magnetic field measurement are widely used in MHD research and other magnetically confined plasmas researches[6]. LM detectors has been updated in all the tokamak devices around the world according to their own experimental requirements [7]. In J-TEXT there are two kinds of LM detectors, which are saddle loops and Mirnov probes placed on the low field side (LFS) and high field side (HFS), respectively [8].

The $n = 0$ pick-up $b_r^{n=0}$ need to be extracted and subtracted to calculate the amplitude and phase of the LM. The widely used method to calculate the magnetic component of odd toroidal mode numbers ($n=1, 3, 5…$) is using two detectors in 180 degrees to subtract the $n=0$ pick-up. This method has been used in devices including the Joint European Torus (JET) [9], the Tokamak Fusion Test Reactor (TFTR) [10], the Texas Experimental Tokamak (TEXT) [11], and the ASDEX Upgrade (AUG) [12]. The magnetic field measured by the LM detector can be expressed as

$$B_r(\theta,\varphi) = \sum b_r^{n=i} \cos(m\theta + n\varphi + \xi^{n=i}), \quad (1)$$

where $\theta$ and $\varphi$ is the poloidal and toroidal location of the LM detector, $m$ and $n$ is the $m/n$ component, $\xi$ is a constant. If only consider the main component in J-TEXT, which are $n = 0$, 1 and 2, for the detector on the middle plane $\theta = 0$ (The following are all based on this situation), equation (1) can be expressed as

$$B_r(\theta=0,\varphi) = b_r^{n=0} + b_r^{n=1}\cos(\varphi+\xi^{n=1}) + b_r^{n=2}\cos(2\varphi+\xi^{n=2}). \quad (2)$$

The $n=1$ pick-up $b_r^{n=1}$ can be calculated through two LM detectors with $\Delta\varphi = \pi$, which is shown in equation (3)

$$b_r^{n=1}\cos(\varphi+\xi^{n=1}) = \frac{B_r(\varphi) - B_r(\varphi+\pi)}{2}. \quad (3)$$

Then $b_r^{n=1}$ and $\xi^{n=1}$ can be calculated by fitting two pairs of these LM detectors, which is at least 4 LM detectors. However, this method required higher spatial resolution on $n = 2$ measurement. $b_r^{n=2}$ and $\xi^{n=2}$ are expressed as

$$b_r^{n=2}\cos(2\varphi+\xi^{n=2}) = \frac{B_{r1} + B_{r2} - B_{r3} - B_{r4}}{4}, \quad (4)$$

where $B_{r1} = B_r(\varphi)$, $B_{r2} = B_r(\varphi+\pi)$, $B_{r3} = B_r(\varphi+\pi/2)$ and $B_{r4} = B_r(\varphi+3\pi/2)$. Two sets of the LM detectors (8 LM detectors) are required and a set of 4 LM detectors must be 90 degrees apart. In J-TEXT, even the HFS Mirnov probes does not meet such spatial resolution requirements. In future devices (such as ITER or CFETR), the diagnosis location distribution is also very limited. If $b_r^{n=0}$ can be extracted through other method, $b_r^{n=2}$ and $\xi^{n=2}$ can be expressed as

$$b_r^{n=2}\cos(2\varphi+\xi^{n=2}) = \frac{B_r(\varphi) + B_r(\varphi+\pi)}{2} - b_r^{n=0}, \quad (5)$$

the requirement number of LM detectors would be reduced to 4 instead of 8. $b_r^{n=0}$ is mainly contributed by equilibrium magnetic field with the eddy current and other uncertainties. In J-TEXT, a lumped eddy current circuits model has been used to compensate the axisymmetric equilibrium fluxes pick-up by LM detectors [13].

Machine learning, which has been widely applied in tokamaks [14],[15], can be applied for classification and regression. Disruption prediction is a typical classification [16], while regression can also be applied to deal with various problems in tokamaks[17]. Machine learning could build an end-to-end model between the inputs and outputs by training large amounts of data, which could cover more factors that affect the pick-up.

This paper will introduce a new method to extract $b_r^{n=0}$ by LM detectors for locked mode calculation, which structured as follows: Section II introduces the database and the Power Multiple Time Scale (PMTS) Neural Networks (NNs) on J-TEXT; Section III shows the result of the regression models, which can regress the $n=0$ pick-up with little error. This model also applied in a resonant magnetic perturbations (RMPs) penetration experiment to test the performance; Section IV analyzes the advantage of the PMTS approaches, especially on the frequency domain; Finally, a brief summary is given in Section V.

## II. DATABASES AND PMTS MODEL

In J-TEXT, the equilibrium magnetic field are Horizontal magnetic field (HF), Vertical field (VF), Ohm field (OH) and Toroidal field (TF). Ideally, we can achieve four independent pick-up models of the four types of fields by energizing each magnetic field current independently. However, the Toroidal field will be constant while the LM detectors began to acquire and the OH coils cannot pass large currents (similar as discharges) independently in J-TEXT. Hence, only the model of HF and VF could be achieved. However, only achieve HF and VF could not regress $b_r^{n=0}$ totally during discharge.

Different from the disruption prediction, $n=0$ pick-up extraction does not need so much data for training. The disruption prediction requires the model to have sufficient generalization and to deal with various discharge situations. While $n=0$ pick-up extraction is just an offline calculation method, which only care about the discharges around the calculation purpose. In J-TEXT, the most locked mode researches are under within a small span of plasma parameters. Choosing suitable discharges to train a model is enough for locked mode calculation.

### A. Databases of the three models

In this paper, we first use our model on a field-only discharge (no plasma, but coils VF or HF are energized) to predict $b_r^{n=0}$, showing the potential use of our model. Then we applied the model under common discharge situation. The two equilibrium field databases were selected in the discharges while energized themselves independently during 2018 Autumn. (The LM detectors are LFS saddle loops.) The common discharges with no tearing/locking 2/1 or 3/1 mode during 2019 Autumn were selected as the $n=0$ pick-up. (The LM detectors are HFS Mirnov probes.) The sampling rate of all the inputs and outputs are normalized to 10 kHz. The total output samples are 6k (time range from 0s to 0.6s) to 8k (time range from -0.1s to 0.7s) for each shot. The detailed information of the databases is shown in table 1. The test shots account for 30% of total shots, which are independent to the training shots.

### B. The Power Multiple Time Scale (PMTS) Neural Networks (NNs)

A new approach called Power Multiple Time Scale (PMTS), which could input more previous time information to the model, has been come up with in this section. The traditional Neural Networks is regressing the data point to point. As shown in equation (6).

$$y(t) = f(t) \qquad (6)$$

However, the output in a time series, is not only related to the time now, but also related to the time before (just like the $b_r^{n=0}$ extraction).

Multi-Time-Scale (MTS) [18] approach has been applied in disruption prediction by using the input parameters to the model at different time. As shown in equation (7).

$$y(t) = f(t, t-\Delta t_i, t-2\Delta t_i, t-3\Delta t_i) \qquad (7)$$

Where $\Delta t_i$ is hyperparameters not the corresponds to the sampling rate. The subscript $i$ means the number $\Delta t$. In this paper, we chose $\Delta t$ as 0.2ms and 1.5ms. While, this approach did not perform well, especially during the equilibrium field current ramp down (This result will be shown in Section IV). This approach doesn't have the ability to remember more previous times, therefore, this approach has been updated to a more powerful approach. The expression is following

$$y(t) = f(t, t-\Delta t, ..., t-\alpha\Delta t), \qquad (8)$$

where $n$ is also hyperparameter and $1 \le \alpha \le \dfrac{t-t_0}{\Delta t}$. $t_0$ is the first time point. However, this approach contains too much inputs, which will consume a lot of computing resources and easily cause overfitting. Then an approach that can consider more previous time information without occupying too much computing resources and prevent overfitting is proposed.

$$y(t) = f(t, t-1^k\Delta t, ..., t-\alpha^k\Delta t), \qquad (8)$$

where $n$ and $k$ are hyperparameters and $1 \le \alpha \le \sqrt[k]{\dfrac{t-t_0}{\Delta t}}$.

The weakness of PMTS is that the beginning of the output data has to be abandoned cause there is no enough input data. The abandoned data is determined by the hyperparameters $\alpha$, $\Delta t$ and $k$ ($t \ge \alpha^k\Delta t + t_0$) according to the needs of the problem to be solved. In this paper, $\alpha$ was selected as 20, $\Delta t$ was selected as 0.2ms and $k$ was selected as 2. The hyperparameters have not been carefully scanned, however, this combination of hyperparameters is the best performing of the hyperparameters we have tried. The choice of $\Delta t$ could be physics guided. In this paper, we considered that the frequency of change of the equilibrium magnetic field will not be higher than 5kHz, so $\Delta t$ is selected as 0.2ms. Recurrent neural network (RNN), especially Long Short-Term Memory network (LSTM) could also solve this kind of issues [19]. TCN [20] and WaveNet [21], which are based on causal convolution, dilated convolution and residual convolution, also have the ability to solve time series problems. However, their networks are too deep to extract the $n=0$ pick-up, which is complicated and needs more data. PMTS is a more simplified approach for fully-connected neural network. The approach can also be flexibly adjusted according to other physical problems that need to be solved.

The models were selected as fully-connected Neural Networks with two hidden layers as shown in figure1. The features of VF/HF model is their current $I_{HF/VF}$. For the model of $n=0$ pick-up, the features were plasma current ($I_P$), toroidal magnetic field ($B_t$), loop voltage ($v_{loop}$), displacement of $x$ and $y$ axis ($d_x$ & $d_y$) and three equilibrium field (VF, HF and OH) current ($I_{VF}$, $I_{HF}$, $I_{OH}$). The activation function is *Tan-Sigmoid*. The target error is described by the mean square error (MSE).

TABLE I
DATABASES OF THE THREE MODELS.

| Database | No. of Shots | Experimental Period | Device | Detector |
|---|---|---|---|---|
| VF data | 15 | 2018 Autumn (1051879-1054267) | J-TEXT | LFS saddle loops |
| HF data | 20 | 2018 Autumn (1051871-1052597) | J-TEXT | LFS saddle loops |
| $n=0$ pick-up data | 30 | 2019 Autumn (1066010-1066255) | J-TEXT | HFS Mirnov probes |

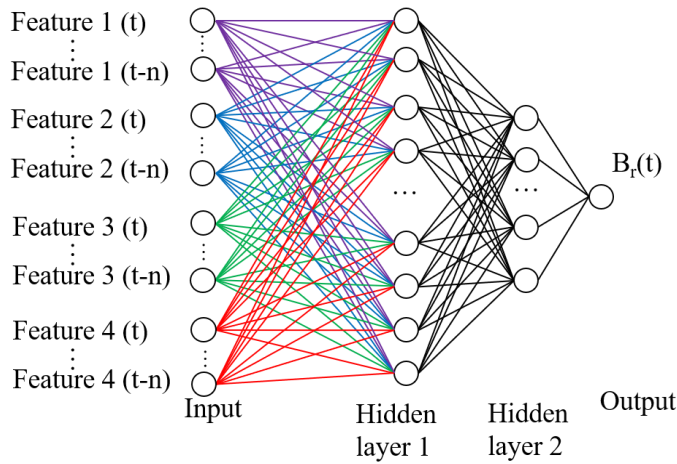

Fig. 1. The Fully-connected Neural Networks with two hidden layers of the three models. The inputs of the models are features at different time.

$R^2$ is a parameter to evaluate the performance of the regression model in machine learning. The closer to 1 the better the model performance. Three models (VF, HF and $n=0$ data during discharges) based on PMTS has been trained with the coefficient of determination $R^2$ equals to 0.98, 0.98 and 0.96, respectively. The performance on test set is 0.96, 0.96 and 0.91. The performance of all these models is shown in table 2.

TABLE 2
PERFORMANCE OF THE THREE MODELS.

| Database | Training set $R^2$ | Test set $R^2$ |
|---|---|---|
| VF data | 0.98 | 0.96 |
| HF data | 0.98 | 0.96 |
| $n=0$ pick-up data | 0.96 | 0.91 |

## III. EXTRACTED RESULTS AND ANALYSIS

In this section, the regressed performance will be shown in three cases: Case A. The test shot result of VF and HF model with similar accuracy to the lumped eddy current circuits model (compensation error less than 2%). This case will show the ability of PMTS to regress $b_r$; Case B. The test shot result of $n=0$ data model with no MHD instabilities with the compensation error less than 8% during the discharge. The maximum compensation error appears when the plasma current has just ramped up to the flat-top. This case shows the outstanding regression of $b_r^{n=0}$; Case C. The test shot result of plasma response to the RMP during the RMP penetration experiment, indicating that the $b_r^{n=0}$ could be subtracted by only one LM detector. This case shows the simple application of this method.

*Case A: $b_r^{n=0}$ when only energize VF or HF*

A typical response of an LFS LM detector to the current of the HF and VF coils is shown in figure 2. In these two shots, with the HF and VF coils energized independently. The linear compensation on the LFS LM detector of the HF and VF current can be obtained by multiplying the current by one constant value, M. The signal of $B_{LFS}$ in Fig. 2 (a) is unlike that of M*$I_{HF}$: $B_{LFS}$ does not have a flat-top, and after $I_{HF}$ is cut off, the measured value is not zero, but experiences a slow decay. This provides that there is eddy currents flowing near the LM detector. The response of VF coils even is mainly caused by the eddy current.

The NNs model of HF and VF data fit the nonlinear eddy current perfectly. Figure 3 shows the regression result of the HF model. The blue line is the real measurement signal by the LFS LM detector; Thick red purple is the regression result; the brown line is the HF current multiplied one constant value. The regression results almost the same as the real measurement signal with the compensation error less than 2%, which is shown in figure 3 (b). Figure 4 shows the regression result of the VF model also with the compensation error less than 2%.

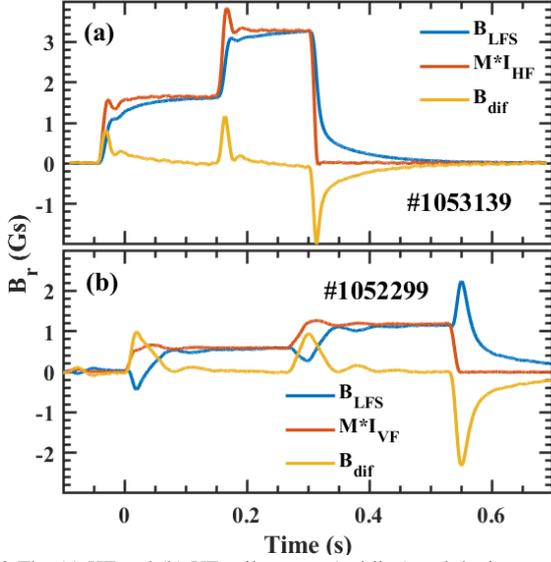

Fig. 2 The (a) HF and (b) VF coil current (red line) and the integrated LFS LM detector signal (blue line) in shot 1053139 and 1052299 with only the HF and VF coils energized.

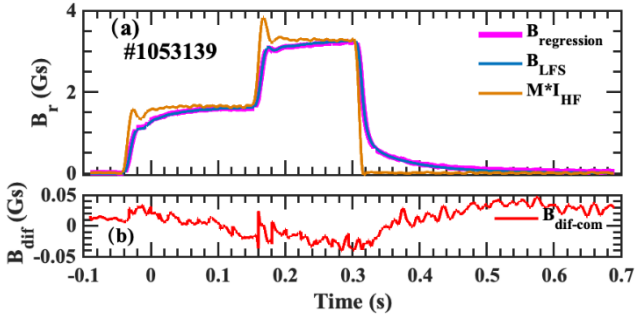

Fig. 3 The HF coil current (brown line), the integrated LFS LM detector signal (blue line) and the regression result (purple line) of the NNs model in shot 1053139 with only the HF coils energized. The compensation error is less than 0.05 Gs (2%).

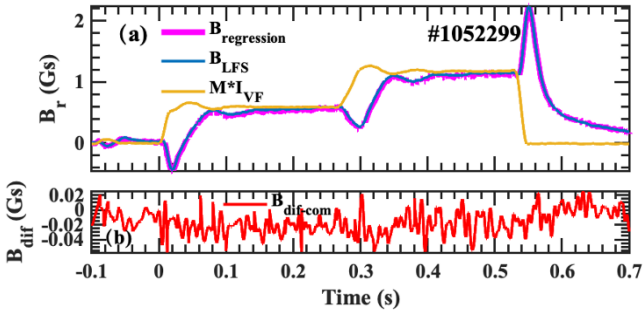

Fig. 4 The VF coil current (brown line), the integrated LFS LM detector signal (blue line) and the regression result (purple line) of the NNs model in shot 1052299 with only the HF coils energized. The compensation error is less than 0.04 Gs (2%).

These two models are single input models, which could also be regarded as a proxy model of the lumped eddy current circuits model. The model learned the eddy current related expressions by itself since no eddy current relative equation in the training approach. The model may be able to learn more complex situations, such as predicting $b_r^{n=0}$ in the real discharge.

*Case B: $b_r^{n=0}$ with no tearing/locking 2/1 mode*

However, the $n = 0$ component during the discharges is much more complicated. The $n = 0$ model described in section II is not a single input model any more. Figure 5 shows a typical discharge (shot #1066279) in J-TEXT with no MHD instabilities, which means only $b_r^{n=0}$ will be measured on the LM detectors. The model of $n = 0$ data is trained through this kind of input signals introduced in section II. The basic plasma parameters in this shot are as follows: $I_p$ = 165 kA, $B_t$ = 1.74 T, edge safety factor $q_a$ = 3.4.

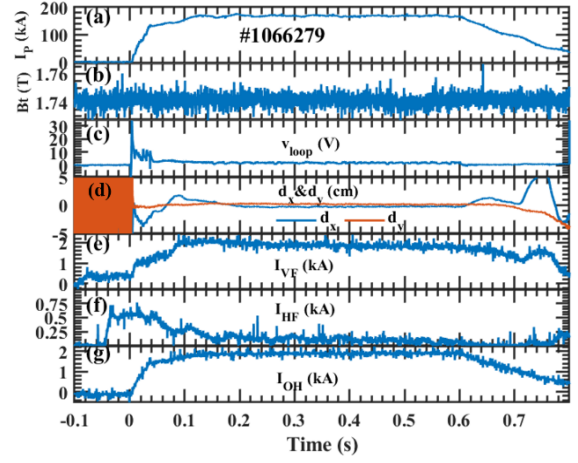

Fig. 5 The input signals of the $n = 0$ model. (a) is plasma current ($I_P$); (b) is toroidal magnetic field ($B_t$); (c) is loop voltage ($v_{loop}$); (d) is displacement of $x$ (blue) and $y$ (brown) axis ($d_x$ & $d_y$); Three equilibrium field (VF, HF and OH) current (IVF, IHF, IOH) are represented by (e), (f) and (g), respectively.

As shown in figure 5, although $B_t$ is almost constant during the discharge, this signal is also selected to extract the $b_r^{n=0}$. The TF coil ramped up from about -1.6s to -0.4s, while the LM detector acquis data at -0.1s, therefore, $B_t$ signal is selected to train the model.

The extraction result in this shot is shown in figure 6. Figure 6 (a) shows the real measurement signal of one HFS LM detector and the extraction result by the $n = 0$ NNs model. The extraction result almost matches on the real signal, even the low frequency perturbation (10~20Hz) on the detector. As described in section II, the weakness of the approach PMTS led to the big mismatch before about 0.02s. Figure 6 (b) shows the compensation error is less than 4 Gs (8%) except for the time before 0.02s. The previous work of measuring locked islands during the RMP penetration has found that the amplitude of magnetic island is about 20~40 Gs. Therefore, the compensation error can be tolerable for further static mode investigations.

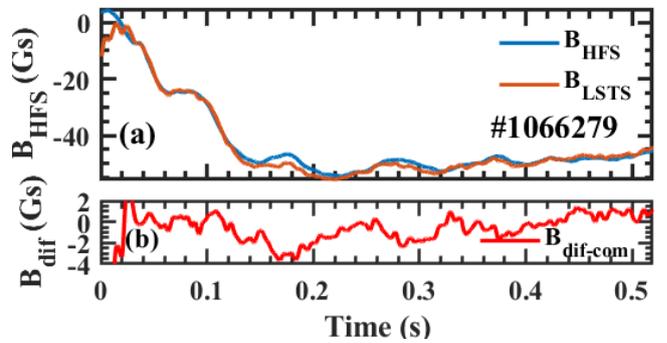

Fig. 6 (a) The real measurement (blue line) of an HFS LM detector and extraction result (brown line). (b) The compensation error is less than 4 Gs (8%).

*Case C: $b_r^{n=0}$ on plasma response to RMP*

The model has been applied in typical J-TEXT experiments, such as RMP penetration experiment. Figure 7 shows a discharge of the RMP penetration experiment in J-

TEXT. The basic plasma parameters in this shot are as follows: $I_p = 160$ kA, $B_t = 1.7$ T, edge safety factor $q_a = 3.3$. As shown in figure 7, the RMP coils were applied from 0.3s to 0.5s. The input signals did not affect by the RMP coils indicated that the RMP coils did not bring uncertainties to the extraction result. The LM detectors also have avoided the influence of magnetic perturbation produced by RMP coils [8]. The model extracted $b_r^{n=0}$, therefore, $b_r^{n>0}$ can be calculated through the difference between the real measurement and extracted $b_r^{n=0}$ (compared with the $n=1$ field generated by RMP, the inherent error field can be ignored).

Figure 8 shows the LM detector signal and model extraction result during the RMP coils applied. Figure 8 (a) is the RMP current signal. Figure 8 (b) shows the real measurement and extraction result. During the RMP applied the extraction result did not predict the plasma response to the RMP. Figure 8 (c) shows the calculation $b_r$ ($\varphi=\varphi_0$) by the PMTS method and traditional method. Through PMTS method, $b_r$ is $n > 0$ pick-up after the extracted $b_r^{n=0}$. While, traditional method represents $b_r^{n=\text{odd}}$. Therefore, the ampliated of $b_r^{n>0}$ could be larger than it of $b_r^{n=\text{odd}}$. Further analysis of the plasma response to the RMP is still on going to figure out what the response real is.

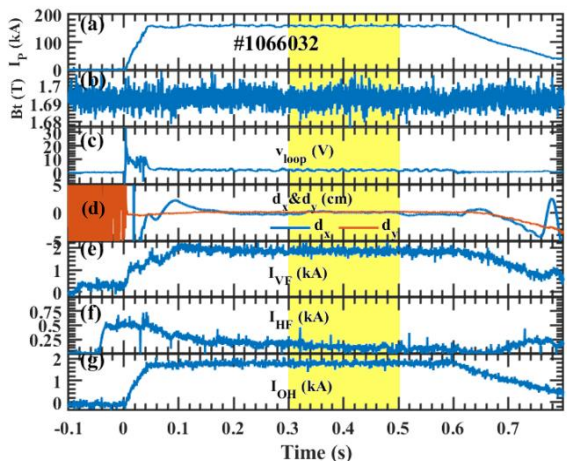

Fig. 7 The input signals of the $n = 0$ model during the RMP penetration experiment. The yellow region applied the 3/1 RMP.

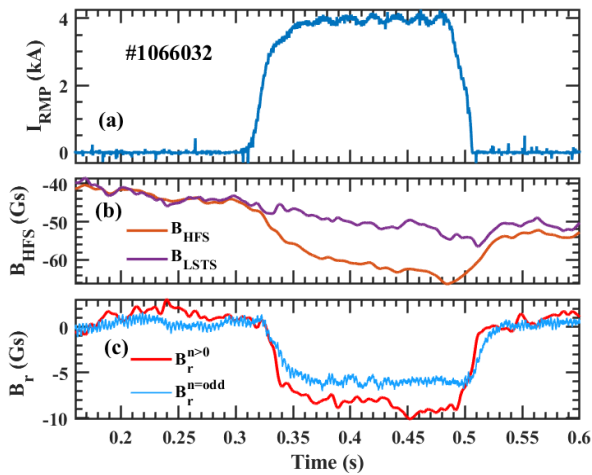

Fig. 8 The measurement and extraction result of plasma response to RMP. (a) RMP current ($I_{RMP}$). (b) The real measurement (purple line) of an HFS LM detector and extracted $b_r^{n=0}$ (brown line). (c) $b_r$ by PMTS method(red) and traditional method(blue).

## IV. FREQUENCY DOMAIN COMPARATION ON TWO APPROACHES

The approach PMTS introduced in section II plays an important role in this work. This section will analyze PMTS in frequency domain and compare from the previous approach MTS. PMTS has a much better performance than MTS, especially in frequency domain.

Figure 9 shows the comparison on two approaches of the HF model both in time domain and frequency domain. Figure 9 (a) and (b) present the comparison in time domain, (c), (d) and (e) present the comparison in frequency domain. In time domain, PMTS has a smaller error than MTS (2% vs. 7%), especially at about 0.4s after $I_{HF}$ cut off. In frequency domain, the performance of PMTS is much better than MTS, the error is almost 4 orders of magnitude smaller. Figure 10 shows the comparison on two approaches of the $n = 0$ model both in time domain and frequency domain. In time domain, PMTS also has a smaller error than MTS (8% vs. 30%), especially when plasma current ramped up. The error of the extraction result on MTS is larger than PMTS when plasma current during the platform. In frequency domain, the performance of PMTS is also better than MTS. However, possibly due to the more complex $n = 0$ model, the error is only one order of magnitude smaller. It is worth noting that the error PMTS is no longer only smaller than MTS in low frequency range, but also in higher frequency range.

As described in section II, PMTS fed more previous time to the model. These previous time will be expressed as the low-frequency component. Therefore, PMTS could do better than MTS especially in low frequency range. The hyperparameters $n$ and $k$ can adjust the time interval of the inputs to prevent overfitting of high frequency components when the interval is too small. This is why in the $n = 0$ model, the accuracy of the PMTS is also higher in the high frequency part than MTS. Meanwhile, there is no high frequency vibrations when plasma current during the platform in the time domain.

In general, PMTS is an approach with outstanding regressing effect in multiple frequency ranges applied on NNs. The more previous time information provides the ability to regress low-frequency data. The hyperparameters $n$ and $k$ provides the adjustable ability to regress high-frequency data.

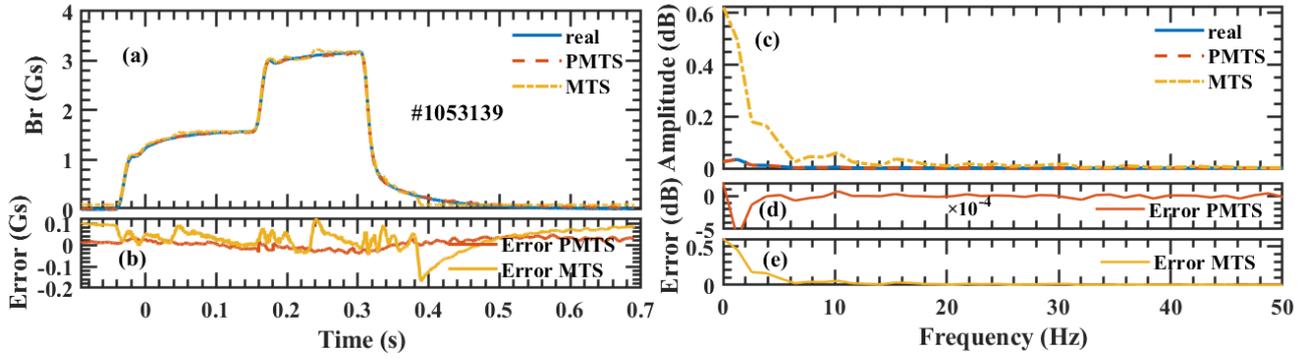

Fig. 9 The comparison on two approaches of the HF model. (a) The real measurement (blue line) of an LFS LM detector, PMTS extraction result (brown line) and MTS extraction result (yellow line); (b) The compensation error of PMTS and MTS in time domain; (c) The FFT analysis of real measurement (blue line), PMTS extraction result (brown line) and MTS extraction result (yellow line); The error of (d) PMTS and (e) MTS in frequency domain.

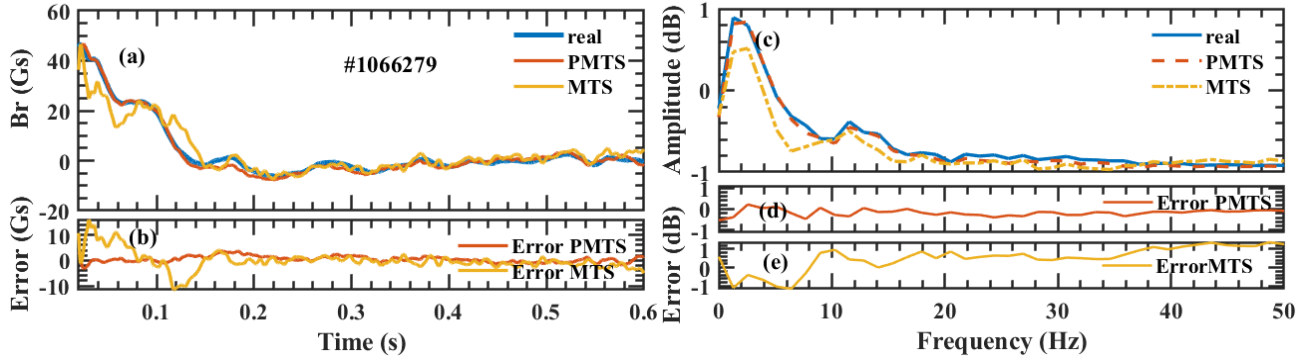

Fig. 10 The comparison on two approaches of the $n = 0$ model. (a) The real measurement (blue line) of an LFS LM detector, PMTS extraction result (brown line) and MTS extraction result (yellow line); (b) The compensation error of PMTS and MTS in time domain; (c) The FFT analysis of real measurement (blue line), PMTS extraction result (brown line) and MTS extraction result (yellow line); The error of (d) PMTS and (e) MTS in frequency domain

## V. SUMMARY

This paper put forward a new method to subtract the $n = 0$ pick-up on the LM detectors based on Neural Networks (NNs) in J-TEXT. An approach called Power Multiple Time Scale (PMTS) has been developed with outstanding regressing effect in multiple frequency ranges. Here the input of $d_x$ and $d_y$ also brings plasma vibrations information to the model, which contains more information than the lumped eddy current circuits model in J-TEXT before. Two single input model for VF and HF data and a multiple inputs model for $n = 0$ data in real discharges have been set up. The compensation error of single input model is less than 2%. While, the compensation error of multiple inputs model is less than 8%. The $n>0$ pick-up generated by RMPs can be obtained after subtracting the extracted $b_r^{n=0}$ by only one LM detector, which makes the equation (5) come true. Therefore, the $n = 2$ LM measurement will only need 2 sets of LM detectors. Each set needs one pair of LM detectors 180 degree apart. In J-TEXT current situation, HFS LM detectors have such situations to carry out future $n = 2$ LM-related work. The LM detectors has a more fixable distribution for the LM measurement. This method can also optimize the LM distribution because of the more flexible choices.

The approach PMTS can be applied to any other application scenarios where the data frequency range covers a wide range. The $n = 2$ LM measurement will be carried out in J-TEXT in the next step.

## ACKNOWLEDGEMENT

Thanks for the suggestions provided in the private communication with Dr. Wei Zheng. This work was supported by National Key R&D Program of China under Grant (No. 2019YFE03010004) and by National Natural Science Foundation of China (NSFC) under Project Numbers Grant (No. 12075096 and No. 51821005).

## DATA AVAILABILITY

The data that support the findings of this study are available from the corresponding author upon reasonable request.